\documentclass[twocolumn,showpacs,preprintnumbers,amsmath,amssymb]{revtex4}
\usepackage{graphicx}% Include figure files
\usepackage{dcolumn}% Align table columns on decimal point
\usepackage{bm}% bold math

\begin{document}
\bibliographystyle{prsty}

\title{Critical role of pinning defects in scroll-wave breakup in
%disordered 
active media}
%\title{Inexcitable obstacles can induce
%transition to scroll wave break-up in three-dimensional active
%media}

\author{S. Sridhar$^1$, Antina Ghosh$^2$ and Sitabhra Sinha$^1$}
\affiliation{$^1$ The Institute of Mathematical Sciences, CIT Campus, Taramani, Chennai 600113, India.}
\affiliation{$^2$ Van der Waals Zeeman Institute, University of Amsterdam, Valckenierstraat 67, 1018 XE Amsterdam, The Netherlands.}

\date{\today}

\begin{abstract}
The breakup of rotating scroll waves
%a characteristic dynamical pattern of activity seen 
in three-dimensional excitable media has been linked to
important biological processes. The known mechanisms for this
transition almost exclusively involve the dynamics of the scroll
filament, i.e., the line connecting the phase singularities.
%Spiral waves and their generalization in three-dimensional excitable media,
%scroll waves,
%are of interest not only from the point of view of the physics of 
%pattern formation, but also for their role in diseases of the heart.
%While the different routes for spiral waves breaking up giving rise to 
%spatiotemporal chaos has been well-studied, so far the only known transition
%mechanism to chaos for scroll waves has been known to be filament breakup
%at the core region. 
In this paper, we describe a novel defect-induced route to breakup of
a scroll wave pinned by an inexcitable obstacle partially extending
through the bulk of the medium.
%induces helical winding of the wave around it. 
The wave is helically wound around the defect inducing sudden changes
in velocity
components of the wavefront at the obstacle boundary. This results in
breakup far from the filament, eventually giving rise to
spatiotemporal chaos. Our results suggest a potentially critical
role of pinning obstacles in
the onset of life-threatening disturbances of cardiac activity.
%does not occur in the absence of the obstacle. 
%The initial wave-break
%leading eventually to chaos occurs far from the filament
%at the boundary with the obstacle,
%where a transition from a quasi-two dimensional front to a
%three dimensional spherical front occurs. 
\end{abstract}
%\pacs{05.45.-a, 87.18.Hf, 87.19.Hh}

\maketitle
%In Chapters~\ref{chapter2} and \ref{chapter3} we have primarily
%focused on reentrant wave patterns in two-dimensional media that are
%manifested as spiral waves. However, most excitable systems that we
%encounter in reality are three-dimensional and we have to consider the
%generalization of a spiral wave for such systems, namely, the scroll
%wave~\cite{Winfree87}. The additional dimension
%opens up the possibility of encountering new dynamical phenomena, a
%specific aspect of which
%we will explore in this chapter.
%In particular, we shall focus on the transition to spatiotemporal
%chaos from a scroll wave through a mechanism that is not encountered
%in two-dimensional media.

\begin{figure}
\includegraphics[width=0.98\linewidth, clip]{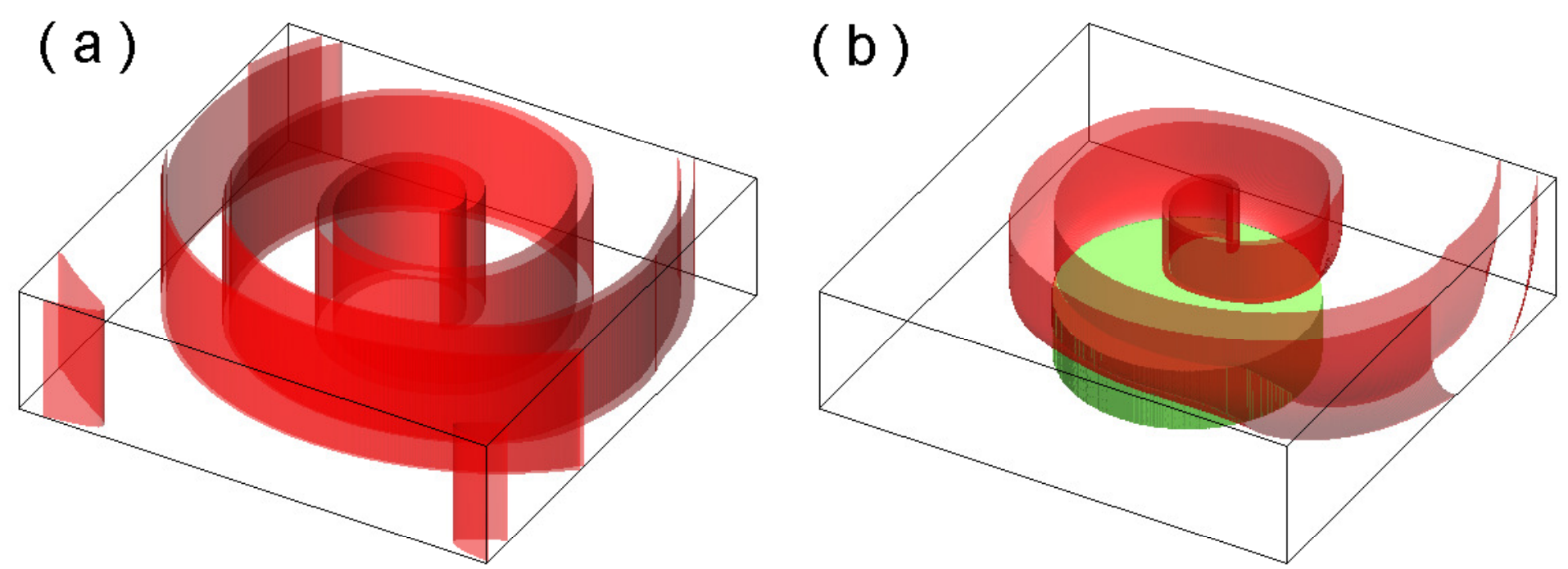}
\includegraphics[width=0.98\linewidth, clip]{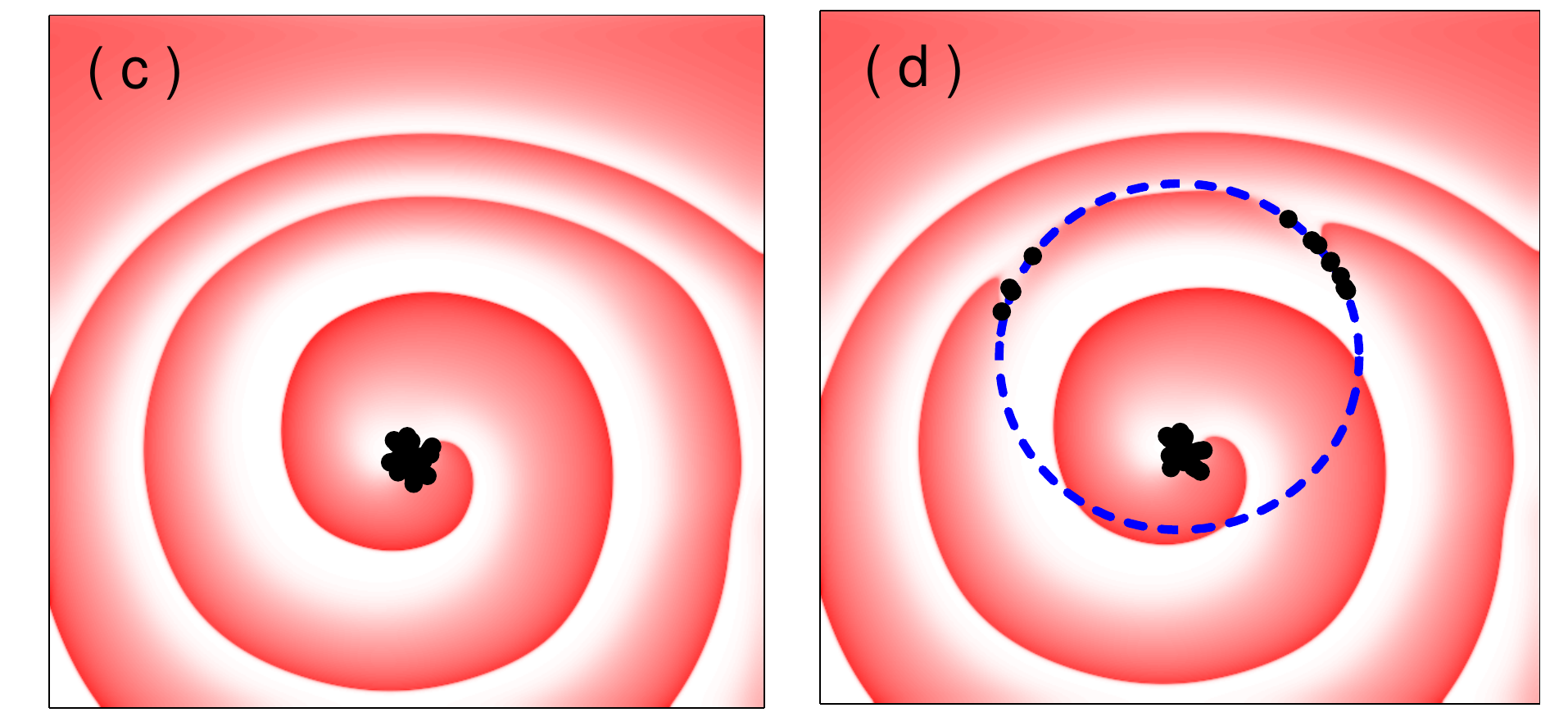}
\caption{Scroll waves in (a) a homogeneous medium and (b) a medium having
an inexcitable cylinder-shaped obstacle that extends only partly
through the bulk of the medium. (c-d) Pseudocolor images of spiral waves
observed in a cross-sectional plane perpendicular to 
the axis of rotation,
for the systems shown in (a) and (b) respectively. 
The black dots correspond to the intersection of the plane with the
filament at different times, indicating the trajectory of the
phase singularity in the plane.
In (d), the plane chosen is just above the upper
bounding surface of the obstacle whose circumference is indicated
using broken lines. The appearance of additional black dots along the
broken circle in (d) indicates a wave-break 
induced by the obstacle at its boundary,
far from the existing scroll filament.
}
\label{Fig1} 
\end{figure}

Spatiotemporal patterns such as rotating spiral waves are
frequently observed in excitable media models that describe
the dynamics of a wide range of physical, chemical and biological
systems~\cite{Cross93,Keener98}. The ubiquity of such waves in
numerous natural processes makes it imperative to understand 
their genesis and the means by which they can be controlled. 
Patterns in homogeneous active media have been the subject of
intense theoretical and experimental investigations for several
decades~\cite{Winfree72,Zykov86,Meron92}. 
However, most natural systems possess significant
inhomogeneities and the dynamics of waves in
{\em disordered} media has come under increasing scrutiny in recent
times~\cite{Panfilov93a,Garfinkel98,Sinha02}.
The heterogeneities considered range from partially or wholly
inexcitable obstacles~\cite{Sinha02,Pumir09}, gradients of excitation
or conduction
properties~\cite{Sridhar10} and anisotropy in the speed of
propagation~\cite{Pumir98}.
%These heterogeneities could be of
%various kinds, a typical examples being a non-conducting or partially
%conducting region. 
Most such studies have focused on
two-dimensional systems and the results show unexpected complexity,
such as fractal basins of attraction for
different dynamical states corresponding to pinned waves,
spatiotemporal chaos and complete termination of
activity~\cite{Shajahan03}.
%In certain cases, the two-dimensional system can be
%reduced to an one-dimensional problem by considering only the reentrant
%circuit surrounding the obstacle. 
Nevertheless, these ``planar" models do not completely 
describe real systems which
are necessarily three-dimensional.
Adding an extra dimension is equivalent to
considering the thickness of the system, so that one can in
principle distinguish between phenomena on the surface and that in the
bulk. More importantly, three-dimensional disordered media can exhibit
novel dynamical phenomena that do not appear in lower dimensions.
A frequently occurring
pattern of activity in such systems is the scroll wave, 
which is a higher dimensional
generalization of the spiral wave.[Fig.~\ref{Fig1}~(a)].
It can be visualized as a set of contiguous rotating
spirals whose phase singularities describe a
continuous line (filament) along the rotation axis perpendicular to
the plane  of the spirals~\cite{Winfree87} [Fig.~\ref{Fig1}~(c)].
Scroll waves have been experimentally observed
in many natural systems including chemical waves in the
Belusov-Zhabotinsky reaction~\cite{Winfree73,Welsh83,Pertsov90}, 
patterns of aggregation during {\em Dictyostelium}
morphogenesis~\cite{Steinbock93,Weijer99}
and electrical waves in heart muscles~\cite{Keener98}.
Indeed scroll waves have been implicated in several types of
arrhythmia, i.e., disturbances in the natural
rhythmic activity of the heart that can be potentially fatal.
Under certain conditions these three-dimensional waves can develop
instabilities and break up into multiple scroll fragments.
The ensuing spatiotemporally chaotic state of excitation is associated
with the complete loss of coherent cardiac activity in life-threatening
arrhythmia such as ventricular fibrillation~\cite{Gray98,Witkowski98}.
Therefore a deeper understanding of the mechanisms that lead to
chaotic activity through breakup of
scroll waves is of great practical significance. 

At present almost
all proposed scenarios for scroll wave breaking involve complex filament
dynamics~\cite{Fenton02}. The question of whether there can be other
mechanisms, especially one involving breakup far from the singularities,
is of special significance when considering disordered media. This is
because heterogeneities such as inexcitable
obstacles can anchor rotating waves and stabilize filament dynamics
preventing the usual breakup scenario. While interaction of scroll
waves with inexcitable obstacles have recently been
investigated~\cite{Vinson93,Pertsov94,Jimenez09,Majumdar11}, the existence of
several types of disorder in the systems considered in these studies
do not clearly reveal the exact
mechanisms for wavebreaks that are involved.
%We want to see whether there can be purely obstacle induced breakup -
%i.e., in the absence of obstacle, the scroll does not break
%Having too much biological detail distracts - it is difficult to see
%whether the breakup is induced by obstacle or fiber rotation or
%bidomain effects etc
%This is why we look at a simple yet biologically realistic model
%where the obstacle induced effect can be clearly distinguished
By focusing on a simplified situation of a defect with regular
geometry
%interacting with a single scroll wave, 
we ask whether there can be a purely
obstacle-induced breakup mechanism for scroll waves.
%that are otherwise stable. 
%(i.e., there is no wavebreak
%in the absence of an obstacle).
%which has no analog in two dimensions.
%The presence of significant heterogeneities in most biological systems
%makes it vital to understand the role of such conduction obstacles in
%initiating instabilities in scroll wave dynamics.

In this paper, we have considered an isotropic three-dimensional
system with an inexcitable obstacle of uniform cross-section that
does not span the entire thickness of the medium [Fig.~\ref{Fig1}~(b)].
This is motivated by the physiological observation that an inexcitable
obstacle may be located deep in the bulk of heart tissue, where it
cannot be detected through electrophysiological imaging of the
surface~\cite{Peters97}.
%Thus, observing from the top, one may not be able to distinguish a
%freely rotating spiral from one
%that is actually anchored deep inside the bulk to an obstacle.
%However the dynamics of such a partially attached
%scroll wave can be distinct from both a pinned spiral wave
%and a freely rotating scroll wave. 
We show a novel dynamical transition that
can occur in such a situation which does not appear in the absence of
a defect, nor in the
effectively two-dimensional situation when the obstacle spans the
thickness of the medium. The differential rotation periods
in various sections of the wave results in a helical structure wound around
the obstacle that in most cases attains a steady state after some initial
transient activity. However, under certain circumstances, the wave can break
near the bounding edge of the obstacle and
far from the scroll wave filament [Fig.~\ref{Fig1}~(d)].
%%%%new insert
%In this paper we show that at the interface
%where the additional velocity components appear, the wave may
%suddenly slow down. Under appropriate conditions, the succeeding wave
%may interact with the refractory tail of the preceding wave resulting
%in a conduction block close to the edge of the obstacle. This causes a
%``tearing" of the front of the succeeding wave which evolves into a
%new scroll filament. This process can occur repeatedly, creating
%multiple filaments that can interact with each other leading to fully
%developed spatiotemporal chaos.
%%%%%%%
This happens when the wavefront velocity decreases sharply at the edge as a
result of sudden change in its curvature. As a result, the interaction
between its waveback and the succeeding wavefront may lead to a
conduction block, ``tearing" the latter which evolves into a pair of
counter-rotating scroll waves. 
The process can repeat, thereby creating multiple filaments which
interact with each other leading to fully developed spatiotemporal
chaos. 
The novelty of the mechanism presented in this paper lies in the fact
that the breakup does not involve the filament, unlike the
previously proposed pathways for the onset of chaos in
three-dimensional excitable media.
It is especially relevant for disordered media as 
the transition to chaos is
essentially defect induced because the wave is
{\em stable} (i.e., does not break or result in the generation of
additional filaments) in the absence
of the inexcitable obstacle. Moreover, this is a genuine three-dimensional
phenomenon that cannot be observed in lower-dimensional systems.

%MODEL DESCRIPTION
%For example, while the rotation period
%of the free spiral wave is essentially limited only by the refractory
%period of the medium, that of the pinned wave is related to the
%circumference of the obstacle. Thus, by increasing the cross-sectional
%area of the obstacle it is possible to enhance the difference in the
%rotation periods between the pinned and freely rotating sections of
%the scroll wave. As a result, when the wave winds around the obstacle,
%in addition to the velocity components along the plane perpendicular
%to the scroll wave filament, there will be propagation along the axis
%parallel to it. 

%\section{Models and methods}
To simulate spatiotemporal activity in three-dimensional excitable
tissue, we use models having the generic form,
\begin{equation} \label{eq1}
        \frac{\partial V}{\partial t}    = -\frac{I_{ion}(V,g_i)}{C_m} +
	{\nabla} . D {\nabla} V,
\end{equation}
%\begin{equation}
%       \partial g_i/ \partial t    = F(V,g_i),
%\end{equation}
where $V$ is the activation variable, typically, 
the potential difference across a cellular
membrane (measured in mV), $C_m$ (= 1 $\mu$F cm$^{-2}$) is the transmembrane
capacitance, $D$ represents the inhomogeneous intracellular coupling,
$I_{ion}$ ($\mu$A cm$^{-2}$) is
the total current density through ion channels on the
cellular membrane, and $g_i$ describe the dynamics of gating
variables of different ion channels.
The specific functional form for
$I_{ion}$ varies for different biological systems.
For the results reported here we have used the Luo-Rudy I
(LR1) model
that describes the ionic currents in a ventricular cell~\cite{LR91}
with the following modifications. The maximum K$^+$ channel conductance $G_K$
has been increased to $0.705$ mS cm$^{-2}$ to reduce the
action potential duration (APD)
making it comparable to that in the
human ventricle~\cite{Panfilov03}. To avoid spontaneous
scroll breakup in the absence of any obstacle, the slow
inward Ca$^{2+}$ channel conductance $G_{si}$ has been set to $\leq 0.04$ mS
cm$^{-2}$~\cite{note1}. 
We have explicitly verified that our results
are not sensitively dependent on model-specific details (e.g.,
description of ion channels) by observing similar
effects in the simpler phenomenological model proposed by
Panfilov (PV)~\cite{Panfilov98}.

The equations are solved using a forward-Euler scheme with time-step
$\delta t = 0.01$ ms in a three-dimensional simulation domain        
having $L \times L \times L$ points. For most results reported here
$L=400$, although we have verified their size independence
by repeating our simulations with different $L$ (viz., upto 600).
The space step used is $\delta x$ = 0.0225 cm, with a standard 
7-point stencil for the Laplacian describing the
spatial coupling.
No-flux boundary conditions are applied on the boundary
planes of the simulation domain as well as along the walls of the
inexcitable obstacle.
The coupling $D$ is set to zero inside the obstacle, while outside it
is $0.001$ cm$^2$s$^{-1}$. 
We have considered obstacles of different shapes, including
cylinders and cuboids, and have not observed qualitative differences
between them. 
Obstacles are characterized by their height $L_z$ (beginning from the
base of the simulation domain and ranging between 0 to $L$) and
cross-sectional area (viz., $\pi R^2$ for cylinders where $R$ is the
radius and $L^\prime \times L^\prime$ for a cuboid).
The initial scroll wave, with its filament aligned along the
$L_z$-axis, is obtained by breaking a
3-dimensional planar wavefront. Qualitatively similar results are
achieved with waves having curved filaments.
%when it arrives at the center of the
%simulation domain starting from one of the boundary planes at 
%$T = 0$. This is achieved by dividing the wavefront into two parts
%along a line parallel to the $L_z$ axis of the obstacle. We then set
%one of the parts to the resting state values, resulting in a broken plane
%front which then dynamically evolves into a rotating scroll wave whose
%filament is parallel to the $L_z$ axis. In the LR1 model simulations,
%an initial plane wave is first allowed to travel through the medium
%to reduce the recovery period. The next wave is then used to create
%the broken plane wavefront, so as to generate a scroll with one full
%turn inside the simulation domain. The location of the
%obstacle at the center of the domain, with its base touching the
%boundary of the simulation domain, ensures that the generated scroll
%wave is attached to it.
%We have also carried out simulations with the initial scroll wave
%being generated by a broken cylindrical front (the axis of the
%cylinder being perpendicular to the $L_z$ axis of the obstacle). While
%the resulting scroll filament is curved compared to that seen in the
%previous method (broken plane wavefront), the subsequent breakup into
%multiple scroll waves is qualitatively similar in both cases.

\begin{figure}
\includegraphics[width=0.95\linewidth, clip]{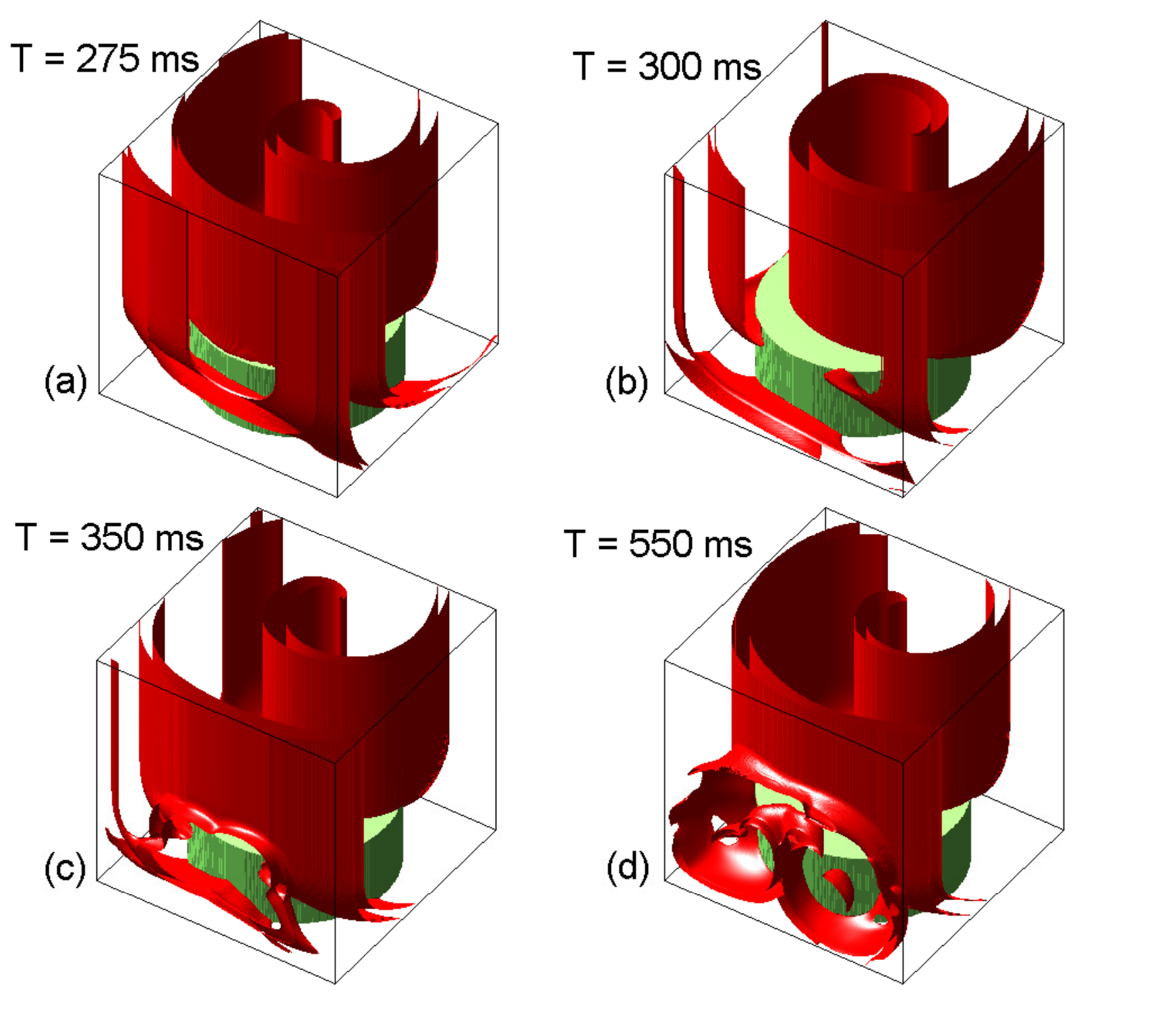}
\caption{Breakup of a scroll wave induced by
a cylindrical obstacle ($R = 3.4$ cm, $L_z = 2.7$ cm)
%The scroll wave is generated by creating a broken wavefront at
%$T = 400$ ms.
(a) By $T=275$ ms after the initial wave-break, the resulting scroll
wave has wound itself around the obstacle. 
%A quasi two-dimensional propagation of the wave along the top surface 
%of the obstacle attains a three-dimensional form 
%when the wave moves past the boundary of the
%top surface of the obstacle
%The curvature of the wavefront changes as it moves past the obstacle.
%and travels down the vertical surface of the obstacle.
(b) A wavefront detaches from the surface of the obstacle,
at the boundary of its upper surface, by $T = 300$ ms. This broken
wavefront 
eventually evolves into new scroll wave filaments (c-d). 
The period of a free (unpinned) scroll wave in this parameter regime
is $T_{free} = 65.13$ ms.
%Interaction between these waves produce multiple excitation wave
%fragments that lead to chaos.
}
\label{Fig2}
\end{figure}

The most important result obtained from our simulations is that a
sufficiently large
pinning defect can promote wavebreaks in an otherwise stable
rotating wave (Fig.~\ref{Fig2}).
%While an extremely small obstacle does not affect the
%dynamics of the scroll wave
At the initial stage, when a broken wave is anchored by the obstacle,
the partial extension of the latter through the bulk of the
medium produces differential rotation speeds of the scroll wave along
the $L_z$-axis. Far above the obstacle the period is closer to $T_{free}$, that
of a free scroll wave unattached to any obstacle, while at
the base of the obstacle the period $T_{pinned} \sim C/c$ is
substantially slower depending upon the circumference $C$ of the obstacle ($c$
being the average propagation speed normal to the wavefront).
%The sequence of events upto the break are as follows.
This results in a helical winding of the scroll wave around the
obstacle until a steady state is achieved when the rotation period
becomes same across the domain [Fig.~\ref{Fig2}~(a)]. The pitch of the
wound scroll wave depends on the obstacle size, with larger $C$
resulting in a tighter winding, i.e., smaller pitch. 
Note that the filament of the resulting scroll wave, which
stretches from the top of the obstacle to the upper boundary of the
simulation domain, remains same as that of a free scroll
wave (Fig.~\ref{Fig1}).
When the wavefront is close to the filament it has velocity components
only along the plane perpendicular to the $L_z$-axis. However, when
the wave crosses the obstacle boundary, it develops a velocity
component parallel to the $L_z$-axis as the front travels down the
vertical sides of the obstacle. The magnitude of the change in
velocity depends on the pitch of the helical winding, and hence on the
circumference of the obstacle, with larger obstacles resulting in greater
velocity differences. There is a corresponding change in the curvature
of the wave along a plane parallel to $L_z$-axis.
As seen in Fig.~\ref{Fig2}~(b), this sudden change in velocity
components as the front crosses the edge of the obstacle upper surface 
may result in detachment of a succeeding wave from the obstacle
surface. 
%as the medium in the neighborhood of the obstacle boundary may
%not have recovered from the passage of preceding waves that underwent 
%a sudden slowing down.
This is manifested as a ``tear" on the wavefront surface as it breaks
%[Fig.~\ref{Fig2}~(c)],
generating a pair of new filaments [Fig.~\ref{Fig2}~(c)].
Further evolution of the system produces more complex wave fragments
as the filaments interact with each other
[Fig.~\ref{Fig2}~(d)], eventually leading to spatiotemporal
chaos.

\begin{figure}
\includegraphics[width=0.95\linewidth,clip]{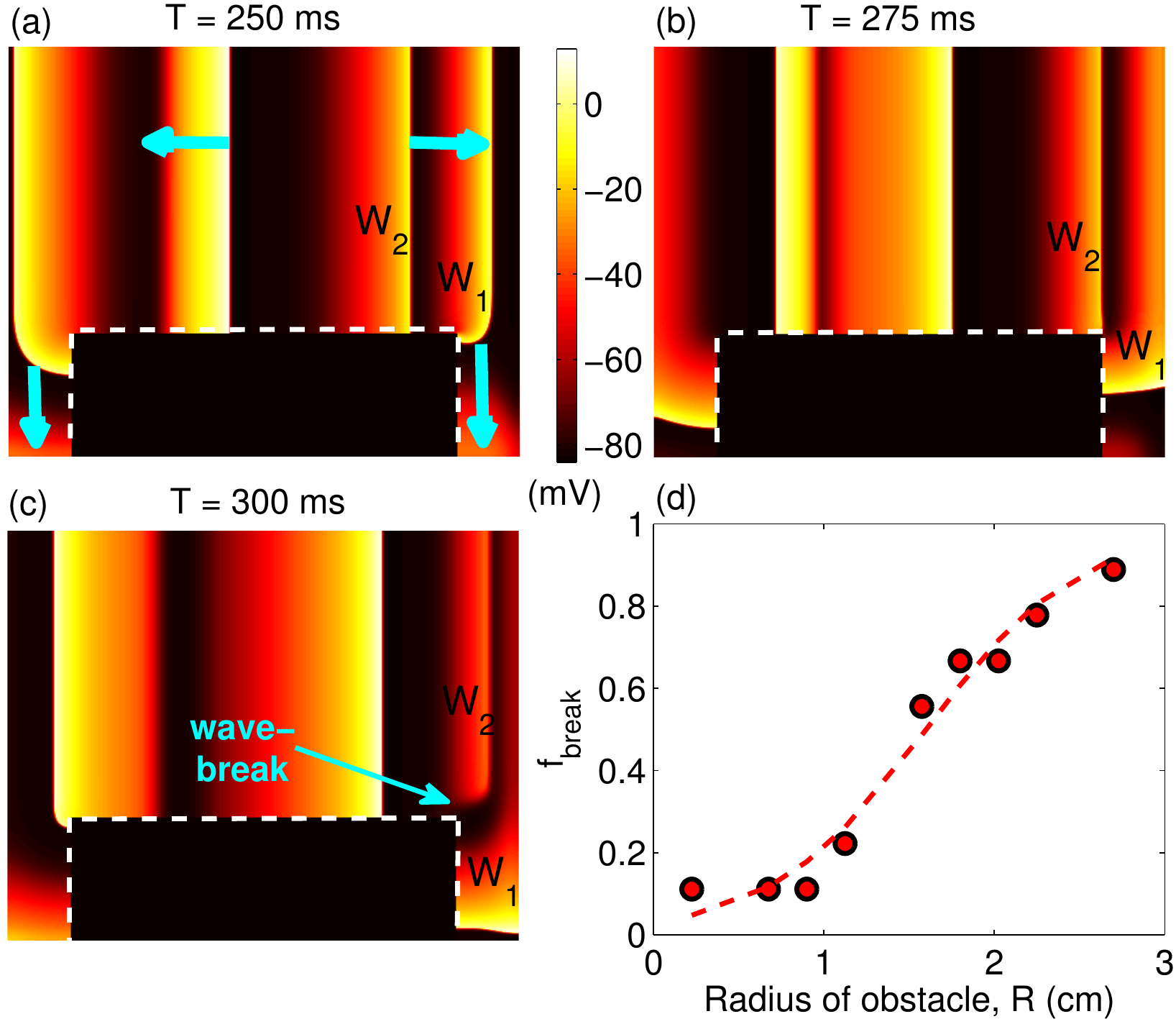}
\caption{(a-c)Pseudocolor plots of the transmembrane potential $V$ 
for a cross-section 
of the system shown in Fig.~\ref{Fig2},
along a plane parallel to the $L_z$-axis.
Boundary of the obstacle (shaded) is marked
by broken lines. The arrows in (a) indicate the direction
of propagation for wavefronts. The wave $W_1$ slows down at the
boundary of the upper surface of
the obstacle.
%as a result of change in the magnitude of the velocity
%components perpendicular and parallel to $L_z$-axis (with a
%corresponding transition from a plane wave to one having convex
%curvature).
%(b) Wave $W_2$, that closely follows $W_1$, encounters
As a result, the wave $W_2$ closely following $W_1$ encounters
a region that has not fully recovered from its prior excitation by
$W_1$, leading to (c) the detachment
of $W_2$ from the surface of the obstacle and formation of a
singularity. (d) The fraction of initial conditions that result in
scroll wave breakup by $T = 3$~s, $f_{breakup}$, as a function of
the obstacle size. The broken line is a sigmoid fit to the data.
}
\label{Fig3}
\end{figure}

To better understand the mechanism by which the pinned scroll wave
breaks, we consider the cross-sectional
view of the
system parallel to the $L_z$-axis [Fig.~\ref{Fig3}~(a-c)]. 
Fig.~\ref{Fig3}~(a) indicates the direction
of propagation as the waves move first along the upper surface
and then down the vertical sides of the obstacle (shaded dark in
the figures). 
The wave $W_1$ slows down as it moves past the boundary of the upper
surface of the obstacle
% [Fig.~\ref{Fig3}~(b)].
This results from the
sudden change in the velocity components of the front mentioned
earlier and is
%the change in the nature of
%its wavefront, from plane to one with a convex curvature. 
associated with an increased curvature of the wave~\cite{Mikhailov94}.
%results in a decrease in its conduction speed~\cite{Mikhailov94}. 
Thus, the wave $W_2$
closely following $W_1$ encounters an incompletely recovered
region at the edge of the obstacle [Fig.~\ref{Fig3}~(b)]. The
resulting propagation block of $W_2$ in the direction parallel to the
$L_z$-axis (immediately adjacent to the obstacle boundary),
dislodges the wave from
the surface of the obstacle and generates a singularity at the point
where the scroll wave breaks [Fig.~\ref{Fig3}~(c)].
This phenomenon can be further enhanced by filament
meandering that results in Doppler-effect induced changes in the
propagation speed of successive waves.
As previously mentioned, larger obstacles result in sharper
differences in the velocity (and curvature) of the front as it crosses
the obstacle,
that suggests increased probability of wavebreak with increasing $C$.
This is indeed observed in Fig.~\ref{Fig3}~(d) supporting
the mechanism outlined above. 
The significance of the results reported here lies in the fact that 
almost all previously reported mechanisms by which scroll waves break
up leading to spatiotemporal chaos necessarily involves the dynamics of
the filament, e.g., negative filament tension in low excitability
regime or filament twist instabilities in the presence of cardiac
fiber rotation~\cite{Fenton02}. By contrast, the novel transition route to
chaos described in this paper is a result of changes in the wavefront
velocity components at the edge of a pinning obstacle, far from
the existing filament. While there exist a few other mechanisms which
can lead to scroll waves breaking, e.g., decreased cell
coupling, these are not exclusively three-dimensional phenomena and
are known to be involved in two-dimensional spiral wave
breakup~\cite{Fenton02}.
%However, the situation reported here does not correspond to any of
%these above scenarios. As mentioned earlier, the wavebreak occurs far
%from the filament in a medium which, apart from the existence of an
%inexcitable obstacle, is isotropic and homogeneous. 

We stress that the mechanism described here has no two-dimensional
analog. While wavebreaks can be
created through interaction between high-frequency excitation fronts with an
inexcitable obstacle in two-dimensional media~\cite{Panfilov93a}, the
corresponding situation in our three-dimensional system, where the waves
encountering an obstacle originate from a scroll wave filament located 
{\em far} from the obstacle (and not pinned by it), does not result in
wavebreaks. This is because, in this case, the wavefront splits and
rejoins as it travels around the inexcitable obstacle without creation
of any new velocity components. Therefore, unlike the pinned
scroll wave considered earlier,
the wavefront conduction speed is not reduced significantly.

The role played by a three-dimensional obstacle in inducing the
breakup of an 
otherwise stable rotating wave
is extremely pertinent for understanding the genesis of certain
cardiac arrhythmias as an ageing heart
gradually accumulates defects through
increased instances of local tissue necrosis~\cite{Zipes04}.
While obstacles in three-dimensional media that do not extend
through the entire thickness of the system have been considered
earlier, these studies focused on the depinning transition of scroll
waves in the presence of drift inducing parameter
gradients~\cite{Vinson93,Pertsov94}.
%typically focused on whether scroll waves
%pinned by such heterogeneities can be detached as a result of a
%parameter gradient that induces a tendency to
%drift~\cite{Vinson93,Pertsov94}. 
In contrast, we show that such defects can give rise to
complex dynamics including transition to chaos.
%by causing the wave
%to wind around it as a result of differing rotation periods across the
%height of the obstacle.
%Our results may also be used to explain scroll breakup even in
%situations where the wave may undergo helical winding in absence of an
%obstacle, e.g., due to fiber rotation in cardiac
%tissue~\cite{Qu00}.

In conclusion, we have shown that the presence of an inexcitable
obstacle in three-dimensional excitable media can result in scroll wave 
breakup far from the filament through a novel physical mechanism.
The helical winding of wave around a 
pinning defect causes sudden changes in the velocity components of the
wavefront as it crosses the boundary of the obstacle, with an
associated increase in the curvature of the wave in the plane parallel
to the axis of its rotation. The resulting enhanced interaction
between successive waves at the bounding edge of the obstacle increases the
probability of a wavefront detaching from the surface of the obstacle
giving rise to new filaments. These wavebreaks can eventually lead to
spatiotemporal chaos, manifested as fibrillation in the heart.
Thus, our results may have consequences for understanding the critical
role of defects (such as, inexcitable regions of necrotic tissue)
embedded deep inside the bulk of cardiac muscle
in the genesis of life-threatening arrhythmia.

%{\it Acknowledgements:}
This work was supported in part by 
IMSc Complex Systems Project (XI Plan) and IFCPAR Project
3404-4. We thank I.~R. Efimov and R. Singh for helpful discussions.


\begin{thebibliography}{2}
\bibitem{note1}We have explicitly verified that the scroll wave does
not break up
in absence of any obstacle by evolving the system for long durations 
($> 10^4$ ms).
\end{thebibliography}
\end{document}